\input harvmac

\Title{LA-UR-96-827}{A THEOREM ON THE LIGHTEST GLUEBALL STATE}

%For more complicated situations, substitute for {\it either\/} argument:
%\Title{\vbox{\baselineskip16pt\hbox{HUTP-88/A000}\hbox{SLAC-PUB 88-001}
%		\hbox{photocopy at own risk}}}
%{\vbox{\centerline{This title is too long to fit}
%	\vskip2pt\centerline{comfortably on one line*}}}
%   \footnote{}{*optional footnote on title}

\bigskip\centerline{Geoffrey B.
West\footnote{$^\ddagger$}{(gbw@pion.lanl.gov)}}
\centerline{High Energy Physics, T-8, MS B285}
\centerline{Los Alamos National Laboratory}\centerline{Los Alamos, NM
87545}\centerline{U. S. A.}

%if too many authors for abstract on same page, say   \vfill\eject\pageno0

%replace this line by \draft  for preliminary versions
	     %or specify \draftmode at some point

%if you want double-space, use e.g. \baselineskip=20pt plus 2pt minus 2pt
\vskip 1.5in
\centerline{\bf ABSTRACT}
\smallskip
This paper is devoted to proving that, in QCD, the lightest glueball state
must be the scalar with
$J^{PC} = 0^{++}$. The proof relies upon the positivity of the path integral
measure in Euclidean
space and the fact that interpolating fields for all spins can be bounded by
powers of the scalar
glueball operator. The problem presented by the presence of vacuum condensates
is circumvented by
considering the time evolution of the propagators and exploiting the positivity
of the Hamiltonian.

\Date{12/95}

\vfill\eject

In this paper I shall show that, if glueball states exist, then the lightest
one must be the $0^{++}$
scalar. There has recently been a renewed flurry of interest, both experimental
and theoretical, in
these very interesting states and the situation is beginning to clarify
\ref\rlone {For a review of recent experimental results and phenomenological
interpretations,
see N.A. Tornquist "Summary of Gluonium95 and Hadron95 Conferences", University
of Helsinki preprint
HU-SEFT-R-1995-16a, hep-ph/9510256 \semi C. Amsler et al., Phys. Lett. {\bf
B355}, 425 (1995) \semi
C. Amsler and F. Close, Phys. Lett. {\bf B353}, 385 (1995) \semi
V.V. Anisovitch and D.V. Bugg, "Search for Glueballs", St. Petersburg preprint
SPB-TH-74-1994-2016.}%
\nref\rlfour {T.Schaffer and E. V. Shuryak, Phys. Rev. Lett. {\bf 75}, 1707,
(1995).}%
\nref\rlfive {J. Sexton, A. Vaccarino and D. Weingarten, Phys. Rev. Lett. {\bf
75}, 4563, 1995.}%
\nref\rlseven {A. Szczepaniak et al., "Glueball Spectroscopy in a Relativistic
Many Body Approach to Hadron Structure", hep-ph/9511422 \semi
M. Chanowitz and S. Sharpe, Nucl. Phys. {\bf B222}, 211, (1983)}%
--\ref\rlthree {M.Schaden and D. Zwanziger, "Glueball Masses from the Gribov
Horizon: Basic Equations
and Numerical Estimates", New York University preprint NYU-ThPhSZ94-1.}.
In spite of this, the situation still remains unresolved and somewhat ambiguous
so exact results such
as that presented here are of some interest. Much detailed analysis has now
been performed on a large
amount of recent experimental data with the result that a few rather good
candidates have emerged
particularly in the region 1.5-1.7GeV \rlone . Potential, bag \rlseven\ and
instanton gas \rlfour\
models suggest that the lowest state should be a scalar and that its mass
should be in the above
range. All of these models, in spite of having the virtue of incorporating the
correct low energy
physics of QCD, are only effective representations of the full theory, and so
their accuracy is
difficult to evaluate. However, recent lattice simulations of QCD based on an
extensive amount of
data are in general agreement with these model results \rlfive . On the other
hand, estimates from a
field theoretic model \rlthree\ indicate that the $2^{++}$ tensor should be the
lightest state
whereas a QCD sum rule analysis indicates that it should be the $0^{-+}$
pseudoscalar  \ref\rlnine
{S. Narison, Z. Phys. {\bf C26}, 209, (1984) and private communication \semi S.
Narison and G.
Veneziano, Int. J. Mod. Phys {\bf A4, no.11}, 2751, (1989).}. This disagreement
between the QCD sum
rules and the lattice measurements is somewhat surprising since they ought to
be the least model
dependent and therefore the most reliable. However, the lattice simulations use
a quenched, or
valence, approximation, which is not generally believed to be a major source of
error, and the QCD
sum rules have difficulty satisfying a low energy theorem. In any case, as
already stated above, the
claim of this paper is that, regardless of the model or approximation used, the
scalar must be the
lightest glueball state. I shall now show why this must be true.

To begin I shall first review some standard formalism as it applies to scalar
and
pseudoscalar glueballs before generalizing to arbitrary states. These spinless
states can be
described by the operators:  \eqn\thirty{{G(x) =
f_{G}F_{\mu\nu}^a(x)F_a^{\mu\nu}(x)}
{\qquad\hbox{and}\qquad\tilde G(x) = f_{\tilde G}F_{\mu\nu}^a(x)\tilde
F_a^{\mu\nu}(x)}}
where  $\tilde F_a^{\mu\nu}(x) \equiv {1\over 2}
\epsilon^{\mu\nu\alpha\beta}F_{\alpha\beta}(x)$ is
the dual field tensor and $f_{G}$ and $f_{\tilde G}$ are constants. The scalar
correlator
\eqn\thirtytwo{\Gamma({\bf x},t) \equiv \langle 0|T[G({\bf x},t)
G(0)]|0\rangle}
has  a standard path integral representation:
\eqn\thirtythree{\Gamma({\bf x},t) =\int{{\cal D} A_{\mu}^a e^{{i\over4}{\int
F_{\mu\nu}^aF_a^{\mu\nu}d^4x}}}{det(\not{D} +m)}{G({\bf x},t) G(0)}}
A sum over quark flavors is to be understood. By inserting a complete set of
states
$|N\rangle$ this can also be written
\eqn\thirtyfour{\Gamma({\bf x},t) =  \sum_{N}{{| \langle
0|G(0)|N\rangle |}^2 e^{(iE_{N}t - i{\bf p_{N} \cdot  x})}}\theta(t) + (t\to
-t)}
from which a corresponding Kallen-Lehmann representation can be inferred.

A useful subsidiary quantity to consider is (for $t>0$)
$$\eqalignno{Q(t) &{} \equiv \int{d^3x \Gamma({\bf x},t)}
&\eqnn\myeqne\myeqne\cr &{}
          = \sum_{N}{{| \langle 0|G(0)|N \rangle |}^2  \delta
^{(3)}({\bf{p_N}})e^{iM_{N}t}}
                &\eqnn\myeqnf\myeqnf\cr} $$
where $M_N$ is the invariant mass of the state $|N\rangle$. The Euclidean
version of this (effectively
given by taking $t\to i\tau$) implies that, when $\tau\to \infty$,
\eqn\thirtyseven{Q_E(\tau) \equiv Q(i\tau) \approx e^{-M_0 \tau}}
where $M_0$ is the mass of the lightest contributing state. An analogous result
can be derived
for ${\Gamma({\bf x},t)}$ via its Kallen-Lehmann representation where the
exponential decay arises
from the large $\tau$ or $|x|$ behaviour of the free Feynman propagator. There
are a couple of points
worth remarking about this before proceeding. First, in pure QCD, where the
$0^{\pm +}$ glueballs are
expected to be the lightest states in their respective channels, $M_0 = M_G$ or
$M_{\tilde G}$. In the
full theory, however, the lightest states are those of 2 pions and 3 pions,
respectively, and the glueballs become unstable resonances and mix with quark
states. In that case
$M_0 = M_{2\pi}$ or $M_{3\pi}$. On the other hand, in the limit when $\tau$
becomes large, but
remains smaller than $\sim{2M_G/{\Gamma^2_G}}$, where $\Gamma_G$ is the width
of the resonance, it
can be shown that the exponential decay law, eq. \thirtyseven , still remains
valid but with a mass
$M_0$ given by $M_G$ rather than $M_{2\pi}$; (a similar result obviously also
holds for the
pseudoscalar case). The point is that, if there are well-defined narrow
resonant states present in a
particular channel, then they can be sampled by sweeping through an appropiate
range of asymptotic
$\tau$ values where they dominate, since $\tau$ is conjugate to $M_N$
\ref\rlten {This and the
closely related problem of mixing between quark and gluon operators will be
dealt with in a
forthcoming paper. For the purposes of this paper, glueballs are defined as
those states created out
of the vacuum by purely (singlet) gluonic operators. It should be pointed out
that, in full QCD in contrast to the pure gauge theory, the proof presented
here breaks down if there are nearby higher resonances which are very broad
(e.g. widths comparable to their masses) or couple much more  strongly (i.e. by
orders of magnitude) to the fields of eq. \thirty\ than the lowest resonance;
see also C. Michael, Nucl. Phys. {\bf B327}, 515
(1989).}.

The basic inequality that we shall employ is that, in Euclidean space,
\eqn\thirtyeight{{(F_{\mu\nu}^a \pm  \tilde F_a^{\mu\nu})^2  \geq 0 }
\qquad\Rightarrow\qquad
{f_G^{-1}G(x) \geq f_{\tilde G}^{-1}\tilde G(x)}}
Although this inequality holds for classical fields, it can be exploited in the
quantized theory by
using the path integral representation, eq. \thirtythree , in Euclidean space
where the measure is
positive definite. The positivity of the measure has been skillfully used by
Weingarten \ref\rleleven
{D. Weingarten, Phys. Rev. Lett. {\bf 51}, 1830 (1983) \semi E. Witten, {\it
ibid} 2351(1983).} to
prove that in the quark sector the pion must be the lightest state. Here, when
combined with the
inequality \thirtyeight , it immediately leads to the inequalities (valid for
$\tau >0$)
\eqn\forty{{{{f_G^{-2}}{\Gamma_E({\bf x},\tau) }\geq {f^{-2}_{\tilde G}}{\tilde
\Gamma_E({\bf
x},\tau)}}}    \qquad\hbox{and}\qquad
{{f_G^{-2}}{Q_E(\tau) } \geq {f^{-2}_{\tilde G}}{\tilde Q_E(\tau) }}}
By taking $\tau$ large (but $<{2M_G/{\Gamma^2_G}}$) and using \thirtyseven,
the inequality
\eqn\fortytwo{M_G \leq M_{\tilde G}}
easily follows. In pure QCD where these glueballs are isolated singularities,
their widths vanish and
the limit $\tau\to \infty$ can be taken without constraint.

Although this is the result we want, its proof presumes the absence of a vacuum
condensate
${E \equiv \langle 0|G(0)|0\rangle}$.  It is generally believed that $E\not=0$
so the lightest state
contributing to the unitarity sum in eq. \thirtyfour\ is, in fact, the vacuum
in
which case $M_0=0$ and the large $\tau$ behaviour of $\Gamma_E({\bf x},\tau)$
is a constant, $E^2$,
rather than an exponential. Thus, the inequalities \forty \enspace are
trivially satisfied for
asymptotic values of $\tau$ since there is no condensate in the pseudoscalar
channel. To circumvent
this problem it is prudent to consider the time evolution of either $Q(t)$ or
$\Gamma({\bf x},t)$
since this removes the offending condensate contribution. For example, (for
$t>0$)
\eqn\fortythree{{dQ(t) \over {dt}}
          = \sum_{N}{{| \langle 0|G(0)|N \rangle |}^2  \delta
^{(3)}({\bf{p_N}}) iM_{N}e^{iM_{N}t}}}
The vacuum state clearly does not contribute to this so, in Euclidean space,
the large $\tau$
behaviour of $\dot Q_E(\tau)$ is, up to a factor $-M_0$, just that of eq.
\thirtyseven . Now,
(for $t>0$), consider the following:
$$\eqalignno{{\partial \Gamma({\bf x},t) \over {\partial t}} &{}
            = \langle 0|{\partial G({\bf x},t) \over \partial t} G(0)|0\rangle
&\eqnn\myeqna\myeqna\cr
&{}
            = \langle 0|i[H,G({\bf x},t)]G(0)|0\rangle &\eqnn\myeqnc\myeqnc\cr
&{}
= -i\langle 0| G({\bf x},t)HG(0)|0\rangle &\eqnn\myeqnd\myeqnd\cr}
$$
where, in the last step, the condition $H|0\rangle =0$ has been imposed. At the
classical level $H$
is positive definite. We can therefore repeat
our previous argument by working in Euclidean space and combining the
inequalities
\thirtyeight\enspace with a path integral representation for \myeqnd\enspace to
obtain (for $\tau
>0$) the inequalities
\eqn\fortyfive{{{ {f_G^{-2}}{\partial\Gamma_E({\bf x},\tau) \over \partial
\tau}
\geq  {f^{-2}_{\tilde G}}{\partial\tilde\Gamma_E({\bf x},\tau)
\over\partial\tau} }}
\qquad\hbox{and}\qquad
{f_G^{-2}{\dot Q_E(\tau)} \geq {f^{-2}_{\tilde G}{\dot {\tilde Q}_E}(\tau) }}}
The large $\tau$ limit then leads to
\eqn\fortyseven{{{ f_G^{-2}}M_G e^{-M_G\tau} \geq {f^{-2}_{\tilde G}}}M_{\tilde
G}
                                                            e^{-M_{\tilde
G}\tau}}
from which \fortytwo \enspace follows even in the presence of condensates
\ref\rltwenty {Notice
that the final result remains unchanged even if the vacuum energy, $E_0$, is
non-vanishing or, what
is essentially equivalent, if a constant, $H_0$, is added to $H$. In either
case the left-hand-side
of \myeqna , or \fiftytwo , is replaced by $e^{iE_0t}\partial / {\partial
t}[\Gamma({\bf x},t){e^{-iE_0t}}]$ leaving
the right-hand-side unchanged. In the large $\tau$ limit this again leads to
\fortytwo\ but with the
masses now defined relative to $E_0$, as one would expect.}.

Although this argument is essentially correct, it still remains incomplete in
that we need to clarify
the nature of the path integral representation for \myeqnd. The point is that
the Hamiltonian, $H$,
that generates time translations must be expressed in terms of canonical
momentum and co-ordinate
field variables. Thus, a path integral representation for \myeqnd\ must first
be written in
Hamiltonian form; unfortunately, however, the resulting measure is not
necessarily positive definite
even in the Euclidean region. Thus, for our above argument to be valid we need
to show that after
integrating the Hamiltonian form over canonical momenta, the classical
expressions for the operators
can still be used and that, in Euclidean space, the resulting measure of the
Lagrangian form remains
positive. I shall first sketch how this comes about in a quantum mechanical
context before
generalizing to field theory. Some of the subtleties encountered are closely
related to (normal)
ordering problems that arise when dealing with operators which, like $H$,
depend on both $P$, the
canonical momentum and $Q$, the canonical coordinate \ref\rltwelve {See, e.g.,
E. S. Abers and B. W.
Lee, Phys. Rep. {\bf 9C}, 1  (1973) \semi I. W. Mayes and J. S. Dowker, J.
Math. Phys.,{\bf 14, no.
4}, 434 (1971).}.  Consider, first, the quantum mechanical analog of \thirtytwo
:
\eqn\fortyseven{\Gamma(t) \equiv \langle 0|T[G(t) G(0)]|0\rangle}  By analogy
with eq.
\thirty\enspace the operator $G$ is to be considered a function of $Q$. A path
integral
representation for this can be generated using the standard procedure of
dividing up the infinite
time interval into discrete infinitesimal sequences of size $\epsilon$ and, at
each discrete time
$t_n$, say, exploiting the completeness of momentum and coordinate eigenstates
(labelled by $p_n$
and $q_n$, respectively) \rltwelve :
 \eqn\fortynine{\Gamma(t) = {\int {dp_1\over {2\pi}}} {\dots} {\int
{dp_{N+1}\over
                            {2\pi}}}{ \int dq_1 }{\dots} {\int dq_N}
e^{i\sum_{n=1}^{N+1} [p_n(q_n -
                            q_{n-1}) - H(p_n,q_n) \epsilon ]} G(q_k)G(q_l)}
Here, $\epsilon = t_{n+1} -t_n$ and $k$ and $l$ are defined such that $t_k=t$
and $t_l=0$. It is
understood that the limits $N\to\infty$ and $\epsilon\to 0$  are to be taken in
such a way that the
total time interval $N\epsilon\to i\infty$ in order to pick out the ground
state expectation value
\rltwelve . An analogous expression for $d\Gamma(t) /dt$ can be derived in a
similar fashion:
$$\eqalignno{{d\Gamma(t) \over dt} & {} = {\int {dp_1
\over{2\pi}}}{\dots}{\int{dp_{N+1}\over{2\pi}}}
                                         { \int dq_1}{\dots} {\int dq_N}
e^{i\sum_{n=1}^{N+1}
                                        [p_n(q_n - q_{n-1}) -
H(p_n,q_n)\epsilon ]} \cr
                                        &
\qquad\qquad\qquad\qquad\qquad\qquad\qquad\qquad\qquad
G(q_k)H(p_m,q_m)G(q_l)&\eqnn\myeqnf\myeqnf\cr}$$
Here $m$ is defined such that $t_m \subseteq [t_k,t_l]$. This freedom in the
choice of $t_m$ is
simply a reflection of the time-invariance of the operator $H$. If $t_m$ lies
outside of this domain, then the integral will vanish since one can then
``undo'' the path integral
using the completeness relations and move the operator $H$ to later and later
times (or earlier and
earlier ones, as appropiate) until it annihilates on the vacuum. A similar
procedure can be used to
verify that this same expression, eq. \myeqnf , can be derived by directly
differentiating the
right-hand-side of eq. \fortynine\enspace with respect to $t_k(=t)$. This
brings down a factor
$i[H(p_{k+1},q_{k+1}) - H(p_k,q_k)]$ leading to the path integral manifestation
of eq. \myeqnc .
However, depending on the  time ordering, one of these factors of $H$ can be
moved (again using
completeness to ``undo'' the path integral) so that it eventually acts on the
vacuum and vanishes. In
this way, one can verify that the operator manipulations going from eqs.
\myeqna\enspace -
\myeqnd\enspace are faithfully reproduced by the path integral and that the
time evolution of
$\Gamma(t)$ is given by eq. \myeqnf .

Typically, and in particular in QCD, $H$ is quadratic in $P$ so the integrals
over the $p_n$ in
eq. \fortynine\enspace are simple gaussians which straightforwardly lead to the
conventional
Lagrangian form (up to an overall vacuum-vacuum amplitude normalization
constant):
\eqn\fiftyone{\Gamma(t) = { \int dq_1 }{\dots} {\int dq_N}
                             e^{i\sum_{n=1}^{N+1} L(\dot q_n ,q_n) \epsilon }
                             G(q_k)G(q_l)}
Here $\dot {q_n} \equiv (q_n - q_{n-1})/\epsilon$. When elevated to field
theory eq.
\fiftyone\enspace becomes eq. \thirtythree .
A similar set of manipulations can be carried out on eq. \myeqnf . Now,
however, there is an added
complication since $H$ occurs not only in its conventional place in the
exponent of the measure, but
also in the integrand itself. Thus, in addition to the usual gaussian integral,
its counterpart
weighted with ${1\over 2}p_n^2$ is also needed. Upon integration this leads to
a factor
$({1\over 2}{\dot q_n}^2 + H_0)$ where $H_0$ is a (positive) constant $\sim
{\epsilon}^{-2}$. This
background energy can be removed by a redefinition of $H(P,Q)$  which only
changes the
energy of the vacuum, $E_0$, without affecting mass differences \rltwenty .
With this definition,
\myeqnf\enspace becomes
\eqn\fiftytwo{{d\Gamma(t) \over dt} = { \int dq_1 }{\dots}{\int dq_N}
                             e^{i\sum_{n=1}^{N+1} L(\dot q_n ,q_n) \epsilon }
                             G(q_k)H(\dot q_m ,q_m)G(q_l)}
As a check on the manipulations leading to this, notice that it is consistent
with the direct
differentiation of \fiftyone\ which was normalized to the vacuum-vacuum
amplitude. The Euclidean
field theoretic version of \fiftytwo\ can now be used to justify the
inequalities leading to our
central result eq. \fortytwo .

The extension of the above argument to the general case showing that the scalar
must be lighter than
all other glueball states, can now be effected. Introduce an operator,
$T_{\mu\nu\alpha\beta\dots}(x)$, constructed out of a  sufficiently long string
of
$F_{\mu\nu}^a(x)'s$ and $\tilde F_a^{\mu\nu}(x)'s$ that it can, in principle,
create an arbitrary
physical glueball state of a given spin. Generally speaking a
given $T$ once constructed can, of course, create states of many different
spins, depending on the
details of exactly how it is constructed. As a simple example consider the
fourth-rank tensor
\ref\rlsixteen {For simplicity color indices as well the trace operator over
color matrices ensuring
the singlet nature of the states have been suppressed.}
\eqn\fiftythree{T_{\mu\nu\alpha\beta}(x) = F_{\mu\nu}(x)F_{\alpha\beta}(x)}
which creates glueball states with quantum numbers $2^{++}$ and $0^{++}$. Now,
in Euclidean space, the
magnitude of any component of $F_{\mu\nu}^a(x)$, or $\tilde F_a^{\mu\nu}(x)$,
is bounded by the
magnitude of  ${{[F_{\mu\nu}^a(x)F_a^{\mu\nu}(x)]}^{1\over 2}}$. Hence, any
single component of
$T_{\mu\nu\alpha\beta}(x)$ must, up to a constant, be bounded by $G(x)$:
\eqn\fiftyfour{T_{\mu\nu\alpha\beta}(x) \leq  f_{G}^{-1}G(x)}     This
inequality is the analog of
\thirtyeight\enspace and so the same line of reasoning used to exploit that
inequality when proving
\fortytwo\enspace can be used here. Following the same sequence of steps leads
to the conclusion that
$M_G$ must be lighter than the lightest state interpolated by
$T_{\mu\nu\alpha\beta}(x)$, from which
the inequality
\eqn\fiftyfive {M(2^{++})\geq M(0^{++}) \equiv {M_G}}
follows. It is worth pointing out that the pseudoscalar analog of this operator
can be similarly
bounded thereby leading to the inequality  $M(2^{++})\leq M(2^{-+})$. This
argument can be generalized
to an arbitrary $T_{\mu\nu\alpha\beta\dots} (x)$ since, again up to  some
overall constants analogous
to $f_G$ of eq. \thirty , it is bounded by some power $(p)$ of $G(x)$; i.e.,
for any of its
components, $T_{\mu\nu\alpha\beta\dots}(x)\leq {G(x)}^p$ . Now, the  operator
${G(x)}^p$ has the same
quantum numbers as $G(x)$ and so can also serve as an interpolating field for
the creation of the
scalar glueball. The same arguments used to prove that this $0^{++}$ state is
lighter than either the
$0^{+-}$ or the  $2^{++}$ can now be extended to the general case showing that
it must be lighter
than {\it any} state created by {\it any} $T$; in other words, the scalar
glueball must indeed be the
lightest glueball state.

Finally, we make some brief remarks about the conditions under which the bound
is saturated. Clearly
the inequalities \thirtyeight\ become equalities when ${F_{\mu\nu}^a (x) =
\tilde F_{\mu\nu}^a
(x)}$  which is also the condition that minimizes the action and signals the
dominance of pure
non-perturbative instantons. In such a circumstance the $0^{++}$ and $0^{+-}$
will be degenerate.
However, the proof of the mass inequality \fortytwo\ only required
\thirtyeight\ to be valid at
asymptotic values of $|x|$. Thus, the saturation of this bound actually only
rests on the weaker
condition that $F$ be self-dual in the asymptotic region where it must vanish
like a pure gauge
field. Similarly, the saturation of the general inequality showing the scalar
to be the lightest
state occurs when {\it all} components of $F_{\mu\nu}^a(x)$  have the same
functional dependence at
asymptotic values of $|x|$. Although this is a stronger condition than required
by the general
asymptotic self-dual condition, it is, in fact, satisfied by the explicit
single instanton solution.
Thus, the splitting of the levels is determined by how much the asymptotic
behaviour of the
non-perturbative fields differ from those of pure instantons. This therefore
suggests a picture in
which the overall scale of glueball masses is set by non-perturbative effects
driven by instantons
(which produce the confining long-range force) but that the level splittings
are governed by
perturbative phenomena.

This investigation was stimulated by some very enjoyable conversations in
Corsica last Summer with
Glennys Farrar, Stephan Narison and particularly Don Weingarten at Gluonium95.
Whilst working on this
problem I have further benefitted from discussions with Rajan Gupta and, most
especially, with Tanmoy
Battacharya. I would like to thank all of these colleagues for their helpful
interactions and the DOE
for its support.

\listrefs

\end